\documentclass[aip, jap, 10pt, groupedaddress, twocolumn]{revtex4-1}
\usepackage[T1]{fontenc}
\usepackage{bm}
\usepackage{graphicx}
\usepackage{amsmath}
\usepackage{amssymb}
\usepackage{sidecap}
\usepackage{dcolumn}
\usepackage{xcolor, soul}
\usepackage{ifpdf}
\usepackage{sidecap}

\begin{document}
\title{\textbf{Strained graphene Hall bar}}

\author{S. P. Milovanovi\'{c}}\email{slavisa.milovanovic@uantwerpen.be}
\author{F. M. Peeters}\email{francois.peeters@uantwerpen.be}

\affiliation{Departement Fysica, Universiteit Antwerpen \\
Groenenborgerlaan 171, B-2020 Antwerpen, Belgium}
\begin{abstract}
The effects of strain, induced by a Gaussian bump, on the magnetic field dependent \textit{transport} properties of a graphene Hall bar are investigated. The numerical simulations are performed using both classical and quantum mechanical transport theory and we found that both approaches exhibit similar characteristic features. The effects of the Gaussian bump are manifested by a decrease of the bend resistance, $R_B$, around zero-magnetic field and the occurrence of side-peaks in $R_B$. These features are explained as a consequence of bump-assisted scattering of electrons towards different terminals of the Hall bar. Using these features we are able to give an estimate of the size of the bump. Additional oscillations in $R_B$ are found in the quantum description that are due to the population/depopulation of Landau levels. The bump has a minor influence on the Hall resistance even for very high values of the pseudo-magnetic field. When the bump is placed outside the center of the Hall bar valley polarized electrons can be collected in the leads.
\end{abstract}


\date{Antwerp, \today}

\maketitle

\section{Introduction}

Graphene\cite{rgr01} can sustain a large amount of strain. Due to its strong $\mathit{sp^2}$ bonds graphene can stretch up to 25$\%$ of its original size without breaking\cite{rstr01}. Furthermore, those mechanical deformations lead to the generation of pseudo-magnetic fields\cite{rpmf1, rpmf2} (for certain strain profiles) that can exceed \cite{rbb1} 300 T. Previous studies showed that triaxial \cite{rts1, rts2, rts3} strain as well as strain generated by bending \cite{rbg1, rbg2} graphene give rise to a quasi-homogeneous pseudo-magnetic field (PMF). Recently, Zhu \textit{et al.} showed that even an uniaxial strain, although in a graphene ribbon with non-uniform width, can lead to a high quasi-uniform pseudomagnetic field \cite{rus1}. On the other hand, imperfections of the substrate can lead to the formation of bubbles and balloons that generate inhomogeneous pseudo-magnetic fields \cite{rbb1, rab1, rbb2}. Moreover, a scanning tunnelling microscope (STM) probe tip can be used to locally deform graphene membranes \cite{rsus0, rsus1, rsus2}. The competition between the van der Waals force, generated between the graphene sample and the probe tip, and the electrostatic force generated by a back gate leads to a convex/concave bending of the graphene sample. It was shown that the pseudo-magnetic field generated in this manner produces regions with opposite sign of the pseudo-magnetic field\cite{rts2, rsus1, rgb1}.
\begin{figure*}[htbp]
\begin{center}
\includegraphics[width=16cm]{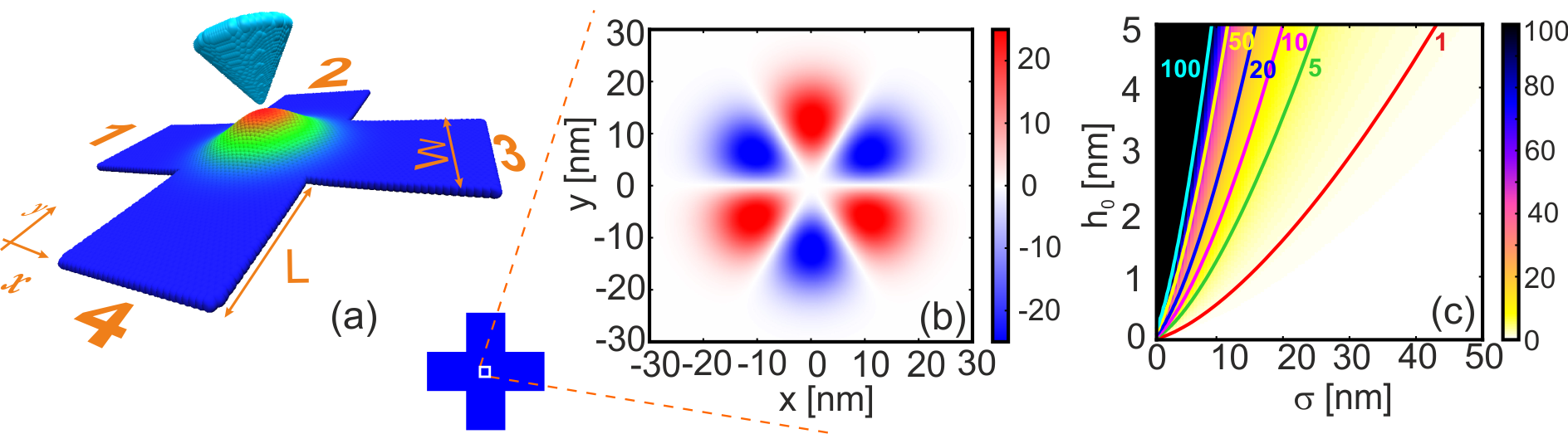}
\caption{(a) Schematic drawing of a Hall bar containing a bump induced by an AFM probing tip. (b) Profile of the pseudo-magnetic field induced by a Gaussian bump with height $h_0 = 3$ nm, and width $\sigma = 10$ nm. (c) Maximal value of PMF, $\left| B_{ps}^{max} \right|$, versus the radius and the height of the bump. The strength of the pseudo-magnetic field i.e. the color scale in (b) and (c) is given in Tesla.}
\label{f1}
\end{center}
\end{figure*}

Strain generated by bumps and other deformations will have an effect on the transport properties because of the generated pseudo-magnetic field\cite{rts1, rgb21, ret1, rgb22, ret2, rgb23, rgf1, rgf10}. Furthermore, imperfections of the substrate are often unavoidable and thus the formation of bumps and bends represents a common problem in the device fabrication process. Up to now the bending/bulging of graphene was investigated by scanning-tunnelling microscopy\cite{rstm01, rstm02} (STM) which measures the effect of the local density of states (LDOS). In a device set-up it would be more convenient to measure the change in the resistance. This motivated us to analyse the electronic response of a graphene Hall bar that is locally deformed by a Gaussian bump situated in the Hall cross. We also applied an external magnetic field and disentangle effects due to the bump from effects of the magnetic field.

In this paper we will present two approaches of transport calculations. First, we present a classical calculation where electrons are described as point-like billiard balls. In the second approach we include the full quantum mechanics of the problem using KWANT \cite{rkw1}, a package for numerical calculations of transport properties based on the scattering of wave functions method. This will allow us to pinpoint which features in the electrical response have a pure quantum mechanical origin and which ones are classical effects.

The paper is organized as follows. In Sec. \ref{sec_meth} we describe the methods that are used for the calculation of the different resistances of interest. The results of our classical calculations are presented in Sec. \ref{sec_cc}. In Sec \ref{sec_tb} the results from the quantum mechanical tight-binding method are presented and examined. Finally, our conclusions are given in Sec. \ref{sec_con}.
\section{Methods}
\label{sec_meth}

Stretching graphene results in changes of the bond length between neighbouring atoms in its lattice. In our nearest neighbour tight-binding model this change results in a modification of the hopping energy given by:
\begin{equation}
\label{e1}
t_{ij} = t_0 e^{-\beta(d_{ij}/a_0 - 1)},
\end{equation}
where $t_{ij}$ is the hopping energy between atoms $i$ and $j$, $t_0 = 2.8$ eV is the equilibrium hopping energy, $\beta = 3.37$ is the strained hopping energy modulation factor, $a_0=0.142$ nm is the length of the unstrained $C-C$ bond, and $d_{ij}$ is the length of the strained bond between atoms $i$ and $j$.

The change of hopping energy is equivalent to the generation of a magnetic vector potential, $\mathbf{A} = (A_x, A_y, 0)$, which can be evaluated around the $\mathbf{K}$ point using\cite{ret1},
\begin{equation}
\label{e2}
A_x - \mathtt{i} A_y = -\frac{1}{ev_F}\sum_j \delta t_{ij} e^{\mathtt{i}\mathbf{K}\cdot \mathbf{r_{ij}}},
\end{equation}
where the sum runs over all neighbouring atoms of atom $i$, $v_F$ is the Fermi velocity, $\delta t_{ij} = (t_{ij} - t_0)$, and $\mathbf{r_{ij}} = \mathbf{r_i} - \mathbf{r_j}$. The pseudo-magnetic field (PMF) is then obtained as
\begin{equation}
\label{e3}
\mathbf{B_{ps}} = \mathbf{\bigtriangledown} \times \mathbf{A} = \left(0,0,\partial_x A_y - \partial_y A_x \right)= \left(0,0,B_{ps}\right).
\end{equation}
Here we used the subscript $"ps"$ to differentiate between the pseudo-magnetic field generated by strain from the applied external magnetic field, $\mathbf{B} = (0, 0, B)$. It is important to mention that the PMF calculated for the  $\textbf{K'}$ point has the opposite direction compared to the one in the $\textbf{K}$ point.

We consider a deformation that has a Gaussian profile given by,
\begin{equation}
\label{e4}
z(x, y) = h_0 e^{-0.5(\mathbf{r} - \mathbf{r_0})^2/\sigma^2},
\end{equation}
with $\mathbf{r} = (x, y)$, and $\mathbf{r_0} = (x_0, y_0)$ is the center of the bump. Such a deformation profile can be realized e.g.  by using the AFM (atomic-force microscopy) or STM probe tip. Due to the van der Waals interaction the probe will attract carbon atoms pulling the graphene upwards, as shown in  Fig. \ref{f1}(a). The parameter $h_0$ in Eq. \eqref{e4} gives us the maximal deformation in the $z$-direction and $\sigma$ is the width of the Gaussian bump. It was previously shown that if the lattice is not relaxed a Gaussian bump induces only a lattice deformation in the $z$-direction\cite{rgb1}. 
\begin{figure}[bp]
\begin{center}
\includegraphics[width=8.5cm]{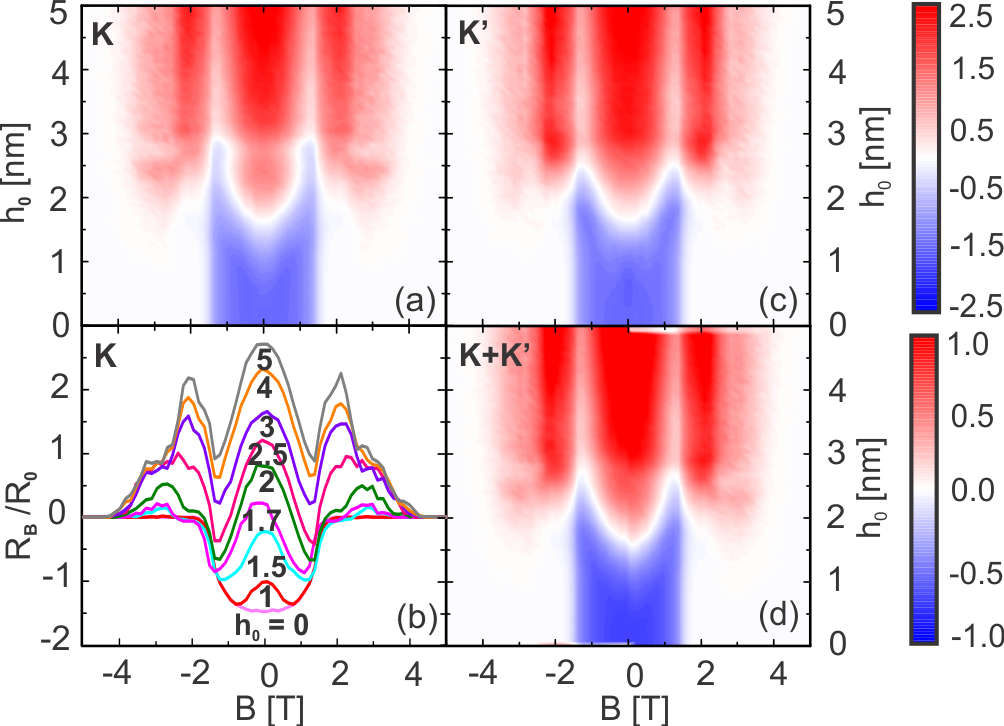}
\caption{(a) Contour plot of the bend resistance $R_B/R_0$ versus the height of the bump, $h_0$, and the external magnetic field $B$ for electrons residing in the $\mathbf{K}$ valley. The radius of the bump was taken to be $\sigma = 10$ nm. The same parameters are used as in Fig. \ref{f1}(b). (b) $R_B$ for different values of $h_0$ i.e. line cuts of (a). (c) The same as (a) but for the $\mathbf{K'}$ valley. (d) The total resistance $R_B$ obtained for electrons from both valleys. The colour scales give the resistance of the contour plots in units of $R_0 = 424.75$ $\Omega$. Maximal value of the PMF is 80 T (for $h_0 = 5$ nm).}
\label{f2}
\end{center}
\end{figure}

The device of interest is a four-terminal, cross-shaped Hall bar, shown in Fig. \ref{f1}(a), with $W = L = 100$ nm. Applying the deformation given by Eq. \eqref{e4} we obtain the induced pseudo-magnetic field as shown in Fig. \ref{f1}(b). The PMF is plotted for $\sigma = 10$ nm and $h_0 = 3$ nm. It exhibits threefold symmetry with alternating positive and negative regions which are along the armchair directions of the graphene lattice while zero field regions are found along the zigzag directions\cite{rgb1}. A contour plot of the maximal PMF versus the height and the radius of the bump is given in Fig. \ref{f1}(c).

First we will perform classical calculations. These simulations are justified for large system sizes, high filling factors, and not too low temperatures. Classical calculations are a valuable tool due to the fact that they describe in general the main features without the fine quantum details and require less computation time. In order to perform these calculations we need an analytical expression for the PMF. This is done in the following way.  Eq. \eqref{e1} can be expanded to first order as,
\begin{equation}
\label{e5}
t_{ij} = t_0(1 - \beta(d_{ij}/a_0 - 1)) = t_0 + \delta t_{ij}.
\end{equation}
Using this expansion we can analytically calculate the PMF following Eqs. \eqref{e2} and \eqref{e3}. The final expression in polar coordinates reads:
\begin{equation}
\label{e6}
B_{ps}(x, y) = c_0\frac{h_0^2}{\sigma^6} r^3 e^{-\frac{r^2}{2\sigma^2}} \sin(3\theta),
\end{equation}
with $c_0 = \hbar\beta / (2ea_0) = 0.781 \times 10^{-5}$  Vs/m. 

Classical calculations for the electrical transport are performed using the billiard model \cite{rbm1, rbm2}. In this numerical model electrons are considered as point particles that are injected uniformly along the width of the different terminals of the Hall bar while the angular distribution of the current carriers follows Knudsen's cosine law. Furthermore, we assume that transport is governed by Newton dynamics. This assumption is valid \cite{rbm2} for $\lambda_F \ll W < l_e$, where $\lambda_F$ is the Fermi wavelength, $W$ is the width of the terminal, and $l_e$ is the mean-free path of the current carriers. This model has been successfully used to describe many classical phenomena in graphene including snake states \cite{rss0, rss1, rss2}, magnetic focusing \cite{rmf01}, different collimation effects \cite{rce01}, etc.

The calculation of the resistance for both classical and the tight-binding method is carried out in the same manner. First, we calculate the transmission coefficients $T_{ij}$ which give us the probability for electrons injected from lead $j$ to be transmitted to lead $i$. These probabilities are then inserted into the Landauer-B\"{u}ttikier formula\cite{rlb1} for multi-terminal structures in order to calculate the current in lead $i$ given by
\begin{equation}
I_i = e/h\left[ (N_i - T_{ii})\mu_i  - \sum_{i \neq j}  T_{ij}\mu_j \right],
\end{equation}
where $N_i, \mu_i$ are the number of modes and electro-chemical potential in lead $i$, respectively. The four-probe resistance is then obtained as $R_{kl, mn} = V_{mn} / I_{kl}$. We are interested in bend resistance $R_{41, 32}$ and Hall resistance $R_{13, 42}$. B\"{u}ttiker \textit showed that in the case of a four-terminal device these resistances can be written as \cite{rlb1},
\begin{equation}
\label{e7}
R_{41, 32} = R_0\frac{T_{34}T_{21} - T_{31}T_{24}}{D},
\end{equation}
and
\begin{equation}
\label{ehr}
R_{31, 42} = R_0\frac{T_{23}T_{41} - T_{43}T_{21}}{D},
\end{equation}
where $D$ is the sub-determinant of the matrix of transmission coefficients given by Eq. (2) in Ref. \onlinecite{rlb1}.
\section{Classical calculations}
\label{sec_cc}
\subsection{Bend resistance}
Figs. \ref{f2}(a-d) show the bend resistance, $R_B = R_{41, 32}$, versus the applied external magnetic field and the height of the bump. The simulations are performed for $\sigma = 10$ nm and $E_F = 0.1$ eV. The resistance is given in units of $R_0 = \frac{h}{2e^2} \frac{\hbar v_F}{|E_F| W}$, which for given set of parameters results in $R_0 = 424.75$ $\Omega$. The results shown in Figs. \ref{f2}(a) and (b) are for PMF calculated in the $\mathbf{K}$ valley, while Fig. \ref{f2}(c) shows the bend resistance obtained when only electrons from $\mathbf{K'}$ valley are considered. The direction of the PMF is opposite in the different valleys and therefore the classical transport simulations have to be performed for each valley separately. The transmission probabilities for both cases are then summed and plugged into the Landauer-B\"{u}ttiker formula. The complete $R_B$ is shown in Fig. \ref{f2}(d). We can see that all three plots show similar features. Therefore, from now on we will consider only electrons from $\mathbf{K}$ valley in order to explain the observed features.
\begin{figure}[htbp]
\begin{center}
\includegraphics[width=8.5cm]{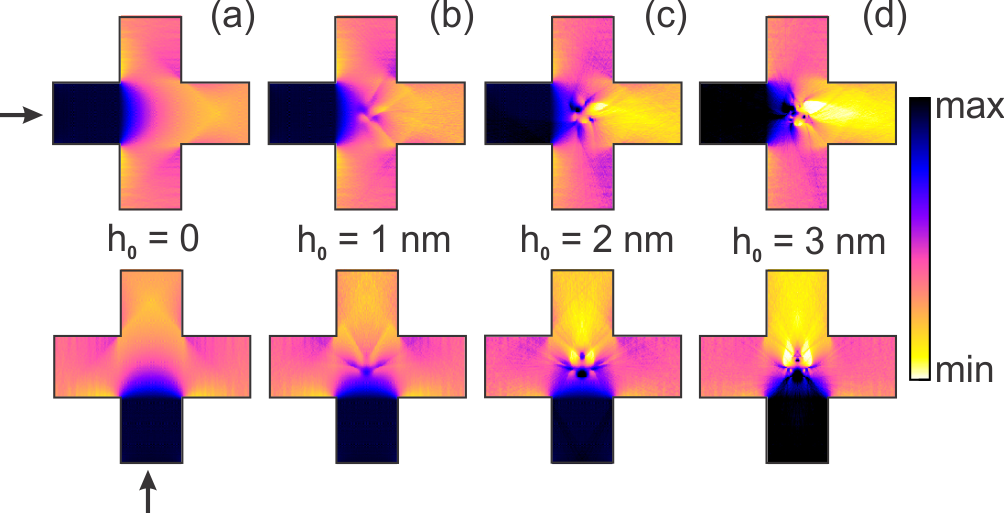}
\caption{Electron current flow in the case of injection from lead 1 (top panels) and lead 4 (bottom panels) for different values of $h_0$ in the absence of an external magnetic field. Values of the PMF are (from left to right): 0, 3.2, 12.7, and 28.7 T.}
\label{f3}
\end{center}
\end{figure}
\begin{figure}[htbp]
\begin{center}
\includegraphics[width=5cm]{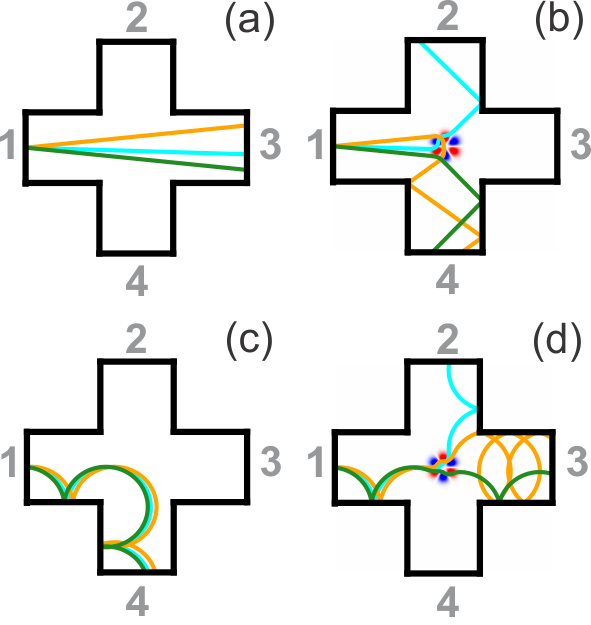}
\caption{Examples of electron trajectories calculated in the case when there is: (a) no pseudo-magnetic nor external field, (b) only PMF, (c) only  external field, and (d) both external and pseudo-magnetic field. The electrons in all figures are injected with the same initial parameters from lead 1.}
\label{f3a}
\end{center}
\end{figure}

To understand the effects of the bump on the transport properties lets investigate how the resistance is changing with increasing $h_0$. For a better visualisation we show cuts of the resistance for different values of the height of the bump in Fig. \ref{f2}(b). The figure shows that the two main effects of the bump are the change of the peak around zero external field and the occurrence of side-peaks.

First, lets try to understand the effect of strain in the case when $B = 0$. The figure shows that by increasing $h_0$, the bend resistance changes from negative to positive. In the ballistic limit negative resistance simply means that the electrons injected from e.g. lead 1 can overpass the central region and flow into the opposite lead (in this case lead 3). This is incorporated in the Landauer-B\"{u}ttiker formula for $R_B$ given by Eq. \eqref{e7}. Hence, if transport towards opposite leads is dominant the bend resistance is negative (second part in the numerator). In the same manner, positive resistance means that the current flows mostly towards the perpendicular leads (in our case leads 2 and 4). This is a consequence of the strain induced PMF that turns the ballistic electrons into the side leads. Similar behaviour was obtained in Ref. \onlinecite{rpil1} for a different system consisting of a ballistic Hall bar made from a GaAs/Al$_{x}$Ga$_{1-x}$As heterostructure with a circular magnetic pillar placed in the center of the Hall bar. It was found that by increasing the magnetization of the pillar the bend resistance changes from negative to positive. 

In Fig. \ref{f3} we show the current flow in the case of electron injection from leads 1 and 4 for different values of $h_0$. These plots are a valuable tool because they show us the effect of strain on the flow of electrons in the device. The system shown in Fig. \ref{f1}(a) has four-fold symmetry, however, the induced magnetic field does not. Hence, injection from lead 1 and lead 4 will differ and both are needed in order to understand the behavior of the resistance. Figures agree with the explanation given above. Indeed, with increasing strain the bump acts as a barrier that blocks the flow of electrons towards lead 3. Notice that for $h_0 = 3$ nm maximal value of the PMF is around 29 T. The top panel of Fig. \ref{f3} shows us the influence of strain on the transmission probabilities $T_{21}$ and $T_{31}$. The flow of electrons from lead 1 towards leads 2 and 4 increases, while the flow towards lead 3 decreases with increasing amplitude of the strain. This effect is clearly seen in Figs. \ref{f3a}(a) and (b) where we plot three electron trajectories for the case when there is no bump and for a bump with $\sigma = 10$ nm and $h_0 = 3$ nm, respectively. Initial conditions for injection are chosen such that the electron interacts with the bump. Figures show that in the former case electrons injected from lead 1 end up in lead 3. However, in the later case i.e. when the bump is introduced into the system, electrons are redirected into the perpendicular leads 2 and 4, and, thus $T_{21}$ and $T_{41}$ increase. Similar conclusions can be made for injection from lead 4, shown in the bottom panel of Fig. \ref{f3}. Transmissions $T_{14}$ and $T_{34}$ increase with increase of $h_0$ while $T_{24}$ decreases. Therefore, an increase of $h_0$ results in an increase of the first term in the numerator of Eq. \eqref{e7} and a decrease of the second term. Consequently, the resistance changes sign - from negative to positive for a particular value of $h_0$.
\begin{figure}[htbp]
\begin{center}
\includegraphics[width=8.5cm]{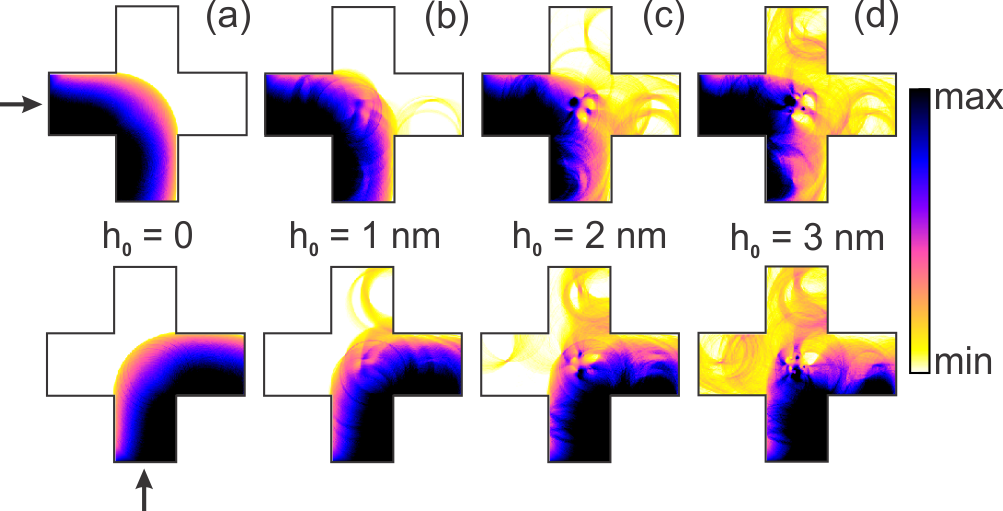}
\caption{The same as Fig. \ref{f3} but in the presence of an external magnetic field $B = 2$ T.}
\label{f4}
\end{center}
\end{figure}

Next, we try to explain the origin of the side-peaks in Figs. \ref{f2}(b).
The figure shows that for $h_0 = 0$, i.e. in the absence of strain, resistance is non-zero only for $|B| \leq 2$ T. For $|B| > 2$ T the cyclotron radius, $r_c$, becomes smaller than (or equal to) the half-width of the terminal and the injected electrons are unable to reach the opposite terminal (see Figs. \ref{f3a}(c) and \ref{f4}(a)). Hence, transmission probabilities $T_{21}$, $T_{31}$, and $T_{24}$ are all zero as well as the bend resistance, according to Eq. \eqref{e7}. When strain is applied the situation changes in a similar way as in the previous case. The electrons are scattered by the bump and dispersed in different directions. However, now the bump serves as a jumping point that enables electrons to overcome the central region and transmit to leads 2 and 3. As a result, the transmission probabilities $T_{21}$, $T_{31}$, and $T_{24}$ all become non-zero and so does the resistance $R_B$. Since more electrons are now transmitted to the perpendicular leads the bend resistance has a positive sign. Furthermore, from Figs. \ref{f4}(b - d) we see that the effect becomes more pronounced with increasing $h_0$ and hence the peak in $R_B$ increases. With further increase of $B$ the bend resistance reduces to zero because of the narrow skipping orbits that transmit all electrons from lead 1 into lead 4 as in Fig. \ref{f4}(a).
\begin{figure}[htbp]
\begin{center}
\includegraphics[width=8.45cm]{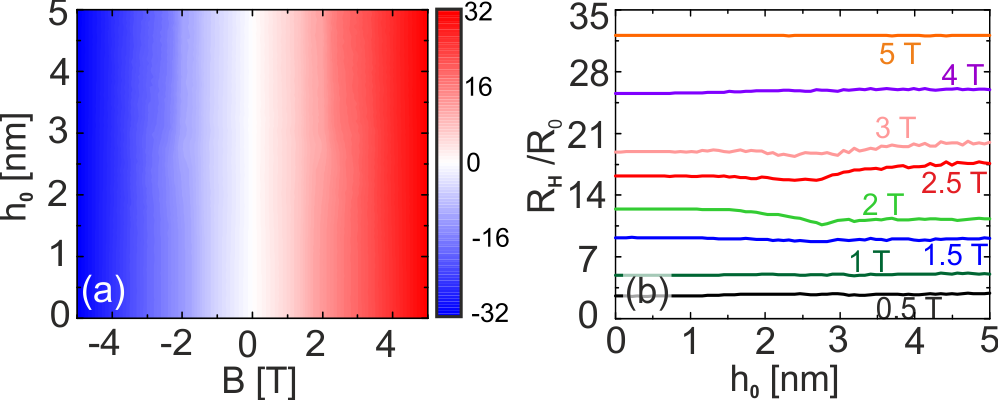}
\caption{(a) The Hall resistance, $R_H = R_{31, 24}$ (in units of $R_0$), versus the applied external field $B$ and the height of the bump, $h_0$, with the bump width $\sigma = 10$ nm. Only electrons from $\mathbf{K}$ valley are considered. (b) Cuts of the resistance versus the height of the bump for different values of applied external field shown in the figure.}
\label{f5}
\end{center}
\end{figure}
\subsection{Hall resistance}

We consider the influence of strain on the Hall resistance given by Eq. \eqref{ehr}. The contour plot of the Hall resistance, $R_H = R_{31, 24}$, is shown in Fig. \ref{f5}(a). The figure shows that the PMF has almost a negligible effect on the Hall resistance. The resistance preserves its linearity with respect to the external magnetic field even for an extremely high values of the PMF ($h_0 = 5$ nm generates PMF of over 50 T). This can be explained in the following manner. In order to have non-zero Hall resistance the current flow has to have a preferential direction, e.g. $R_H$ will be non-zero if there is a higher probability for electrons to end up in one of the perpendicular terminals (2 or 4). At low magnetic fields, the Hall resistance is small since the injected current is well spread over the system (as shown in Fig. \ref{f3}(a)). However, at high fields the current flows towards one terminal only (in the case shown in Fig. \ref{f4}(a) (top panel) that is lead 4). The injected beam is well collimated and hence the Hall resistance is large. When a bump is present the regions with PMF of opposite direction won't induce a significant imbalance in the scattering to the perpendicular terminals since different regions of the bump bend electrons in opposite directions. In Fig. \ref{f5}(b) we show cuts of the Hall resistance versus the height of the bump for several values of $B$. The figure shows that the influence of the bump is visible only for medium fields, i.e. when the cyclotron radius is close to the half-width of the terminals. However, the change of the resistance with $h_0$ is small even at these fields and the maximal absolute difference in the resistance is smaller than $15\%$. Furthermore, at high fields electrons are even unable to reach the bump and consequently their influence on the current flow is further diminished. Similar findings are observed when the electrons from $\mathbf{K'}$ valley are injected. Thus, we can conclude that the PMF generated by the bump placed in the center of the Hall bar affects the Hall resistance only in a small range of magnetic fields and the change of $R_H$ is minor. This result is consistent with Ref. \onlinecite{rbm2} where it was shown that the Hall resistance is proportional to the average magnetic field in the Hall cross.
\begin{figure}[htbp]
\begin{center}
\includegraphics[width=8.5cm]{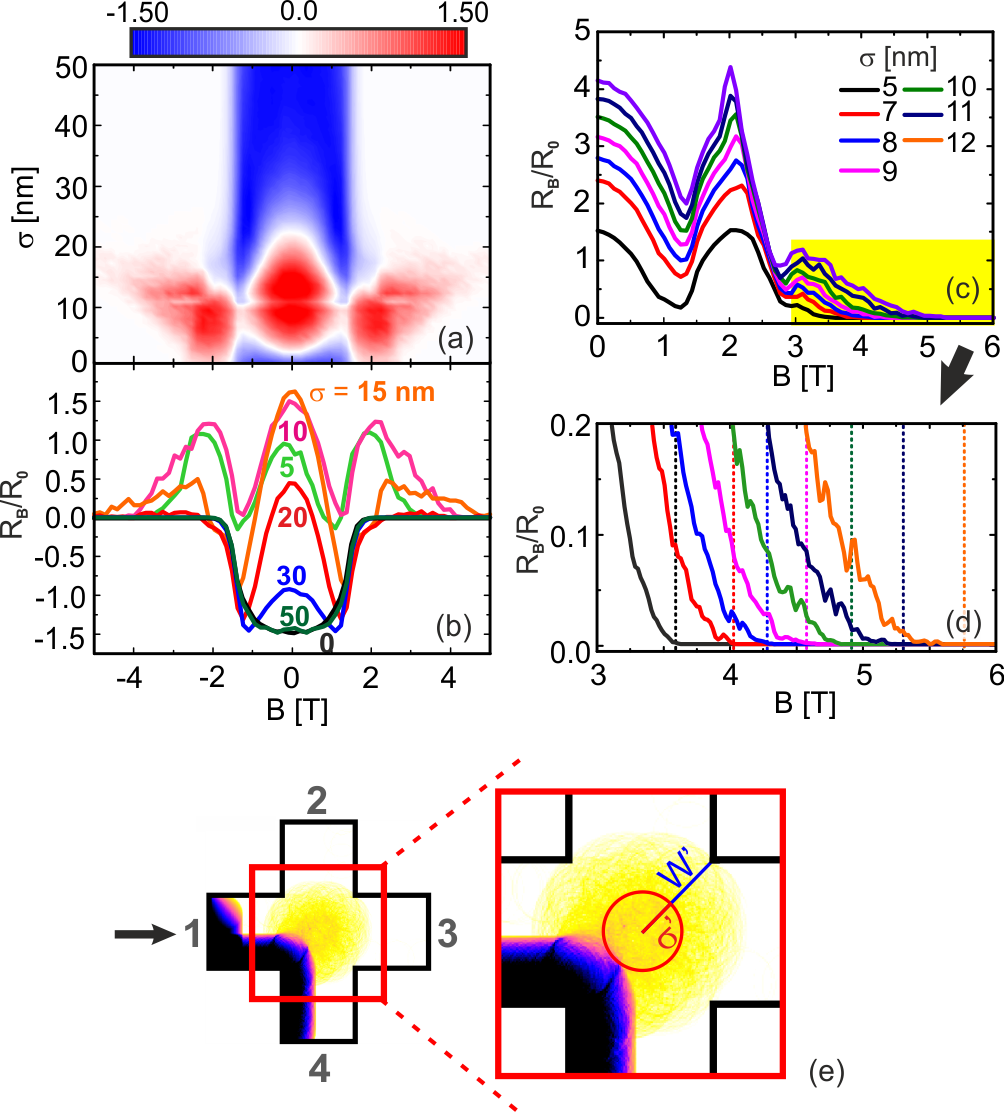}
\caption{(a) The bend resistance versus the applied external magnetic field and the radius of the bump, $\sigma$, for electrons from $\mathbf{K}$ valley and with $h_0= 3$ nm. The color scale is in units of $R_0 = 424.75$ $\Omega$. (b) Cuts of $R_B$ for different values of $\sigma$. (c) The bend resistance for a few values of $\sigma$ and $h_0 = 10$ nm. (d) Zoom of the area enclosed by yellow rectangle from (c). Vertical lines show analytical estimates for the magnetic field where $R_B$ becomes zero. (e) Current density plot for $\sigma = 10$ nm, $h_0 = 10$ nm, and $B = 4.5$ T.}
\label{f6}
\end{center}
\end{figure}
\subsection{Effect of the size of the strained region}

Up to now, we kept the width of the bump fixed while the strength of the PMF was controlled with the height of the bump. The PMF can be also altered by changing the width of the bump, as seen from Fig. \ref{f1}(c). However, with change of $\sigma$ we also control the size of the strained region and thus the amount of electrons affected by it. Hence, in this section we investigate the effect of the width of the bump on our results. Figs. \ref{f6}(a-b) show the bend resistance versus the applied magnetic field and the radius of the bump, $\sigma$. We see that the side-peaks and the positive bend resistance around $B = 0$ occur only in a finite range of $\sigma$ values. For large $\sigma$ the bend resistance approaches the result of the non-strained case. This isn't surprising having in mind that the maximal value of the PMF decreases with increase of $\sigma$, as shown in Fig. \ref{f1}(c). Hence, for $\sigma = 50$ nm  the maximal generated PMF is only $B_{ps}^{max} \approx 0.2$ T. Nonetheless, Figs. \ref{f6}(a-b) confirm that the effects of the PMF on the bend resistance are mirrored by the occurrence of side-peaks and the increase of the resistance around  $B = 0$. 

Notice that the value of magnetic field at which the bend resistance drops to zero, $B_{R_B=0}$, depends on $\sigma$. This is more apparent in Figs. \ref{f6}(c) and (d) where we show the bend resistance for a set of different values of $\sigma$ and a larger $h_0 = 10$ nm. Fig. \ref{f6}(c) shows that the position of the side-peak has a very weak dependence on the width of the bump and occurs for $2r_c \approx W$. This, on the other hand, is not the case with $B_{R_B=0}$ which changes considerably with $\sigma$, as seen from Fig. \ref{f6}(d). As explained previously, when there is no bump the bend resistance is zero for $2r_c(B) < W$, i.e. when the current carriers are no longer able to overpass the central region and transmit to the opposite lead. When the bump is present this will change since now electrons can use the bump to transmit to the opposite lead. To illustrate this situation we plot in Fig. \ref{f6}(e) the current density in a system with a bump using the following parameters: $\sigma = h_0 = 10$ nm ($B_{ps}^{max} = 300$ T) and $B = 4.5$ T. We see that the part of the injected beam is affected by the bump. However, external magnetic field is too strong to bend them towards leads 2 or 3. Hence, this fact can be used to determine $B_{R_B=0}$. We know that $W = W' + \sigma '$, where $W'$ and $\sigma '$ are as in Fig. \ref{f6}(e). $\sigma '$ is the radius of the circle in which PMF is non-zero while $W'$ is the distance from this circle to the nearest point at the edge of the device. Cyclotron radius in graphene is defined as:
\begin{equation}
\label{ecr}
r_c = \left| \frac{E_F}{ev_FB} \right|,
\end{equation}
where $e$ is the elementary charge and $v_F$ is the Fermi velocity.
Using this together with the fact that the resistance drops to zero when $2r_c \leq W'$ we find:
\begin{equation}
\label{ebr0}
B_{R_B=0} = \frac{2E_F}{ev_F(W - \sigma')}.
\end{equation}
The only thing left is to determine $\sigma '$, i.e. the radius at which PMF is non-zero. We use $\sigma ' = 3\sigma$ since we know that the out-of-plane deformation drops to $1 \%$ of its maximum value at $3\sigma$ for Gaussian-like deformation. Solutions of Eq. \eqref{ebr0} for different values of $\sigma$ are shown in Fig. \ref{f6}(d) by vertical lines and notice that they are in good agreement with our numerical results. Nonetheless, one can also notice that the accuracy of the formula decreases with increasing $\sigma$. The reason lies in the estimate of the parameter $\sigma '$.  For larger $\sigma$, the PMF inside radius of $\sigma'$ becomes comparable or lower than the external magnetic field, hence, our solution will deviate more from the actual $B_{R_B=0}$ since we didn't take $B$ into account. Furthermore, this formula allows us to estimate the size of the Gaussian bump. If we know the magnetic field at which the bend resistance drops to zero we can use Eq. \eqref{ebr0} to calculate $\sigma'$.
\begin{figure}[htbp]
\begin{center}
\includegraphics[width=8.5cm]{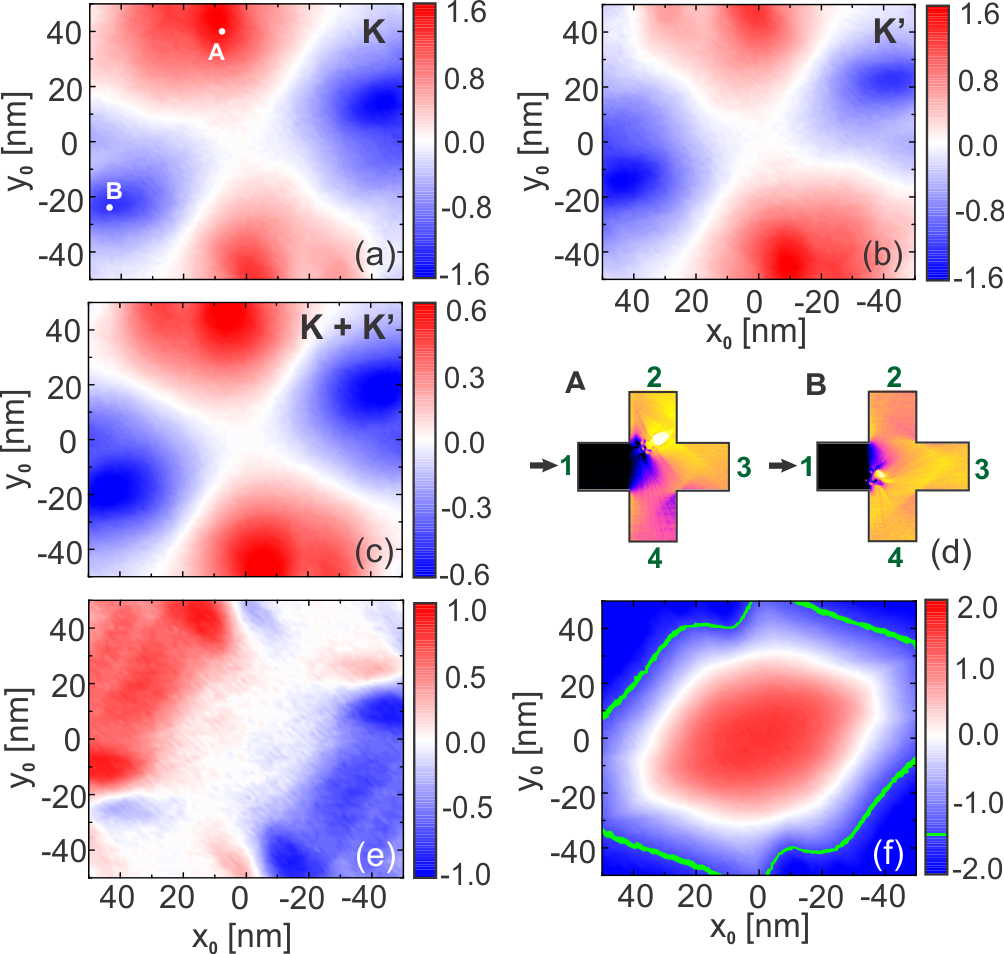}
\caption{The Hall resistance obtained for electrons residing in (a) $\mathbf{K}$, (b) $\mathbf{K'}$ valley, as well as (c) total resistance as function of the position of the center of the bump with respect to the center of the Hall bar. (d) Electron current density plots for the Gaussian bump positioned at $A$ and $B$ from (a). The arrows show the injection lead. (e) The difference between Hall resistances obtained for electrons from opposite valleys, $R_H^{diff}$. (f) The same as (a) but for bend resistance. The simulations are performed for $B = 0$, $\sigma = 10$ nm, and $h_0 = 3$ nm. The color scales are in units of $R_0 = 424.75$ $\Omega$.}
\label{f6a}
\end{center}
\end{figure}
\subsection{Valley polarized current}

As we already pointed out, the Hall resistance is almost unaffected by the presence of the bump. However, this is no longer true if the center of the bump is shifted away from the center of the Hall bar which is a consequence of the fact that the pseudo-magnetic field averaged over the Hall cross is no longer zero \cite{rbm2}. In Figs. \ref{f6a}(a-c) we plot the Hall resistance versus the position of the center of the bump, $(x_0, y_0)$, obtained for electrons residing in $\mathbf{K}$ and $\mathbf{K'}$ valley, and the total resistance, respectively. The simulations are performed for a Gaussian bump with $\sigma = 10$ nm and $h_0 = 3$ nm ($B_{ps}^{max} = 28.7$ T). Similarly as for the bend resistance shown in Figs. \ref{f2}(a-d) these plots show rather quantitative then qualitative  differences. Hence, we focus mainly on the case when only electrons from $\mathbf{K}$ valley are considered. Notice that a non-zero Hall resistance is found although the external magnetic field is set to zero. We see two regions with negative resistance and two regions with positive Hall resistance separated by zero resistance regions (white areas). The explanation for this behavior is rather straightforward and can be seen from the plots of the electron current density, shown in Fig. \ref{f6a}(d). Namely, moving the center of the bump away from the center of the device introduces a preferential direction for scattering of electrons towards one of the perpendicular terminals. In other words, in the case when the bump is in the center of the device approximately the same number of electrons is scattered toward leads 2 and 4. However, when the bump is moved to position A (B) more electrons will be scattered towards lead 4 (2), as shown in the figure. This suggests that the current flowing into lead 4 (2) is valley polarized and hence the Hall resistances for opposite valleys should differ. In Fig. \ref{f6a}(c) we plot the difference between the Hall resistances obtained for electrons residing in different valleys, $R_H^{diff} = R_H^{\mathbf{K}} - R_H^{\mathbf{K'}}$. The figure confirms that for certain $(x_0, y_0)$ values partially valley polarized electrons can be realized in the perpendicular leads.

In Fig. \ref{f6a}(f) we show contour plot of the bend resistance versus the position of the center of the bump. Green curve shows where the resistance equals the resistance without the bump. This plot confirms that the increase of the bend resistance at $B = 0$ observed in Fig. \ref{f2} is even valid when the bump is not positioned at the center of the structure. Furthermore, for a wide range of $x_0$ and $y_0$ the resistance changes its sign from negative to positive.

\section{Quantum mechanical effects}
\label{sec_tb}

In the previous section we obtained the bend and the Hall resistance by using the classical billiard model. Electrons were treated as point particles and the wave nature of the electrons was neglected. In this section we will use the tight-binding model to calculate the resistances using the same parameters as in Section \ref{sec_cc} and we treat the electrons quantum mechanically. In order to suppress strong quantum mechanical effects large filling factors are required and hence the Fermi energy is set to $E_F = 0.3$ eV.
\begin{figure}[htbp]
\begin{center}
\includegraphics[width=7.2cm]{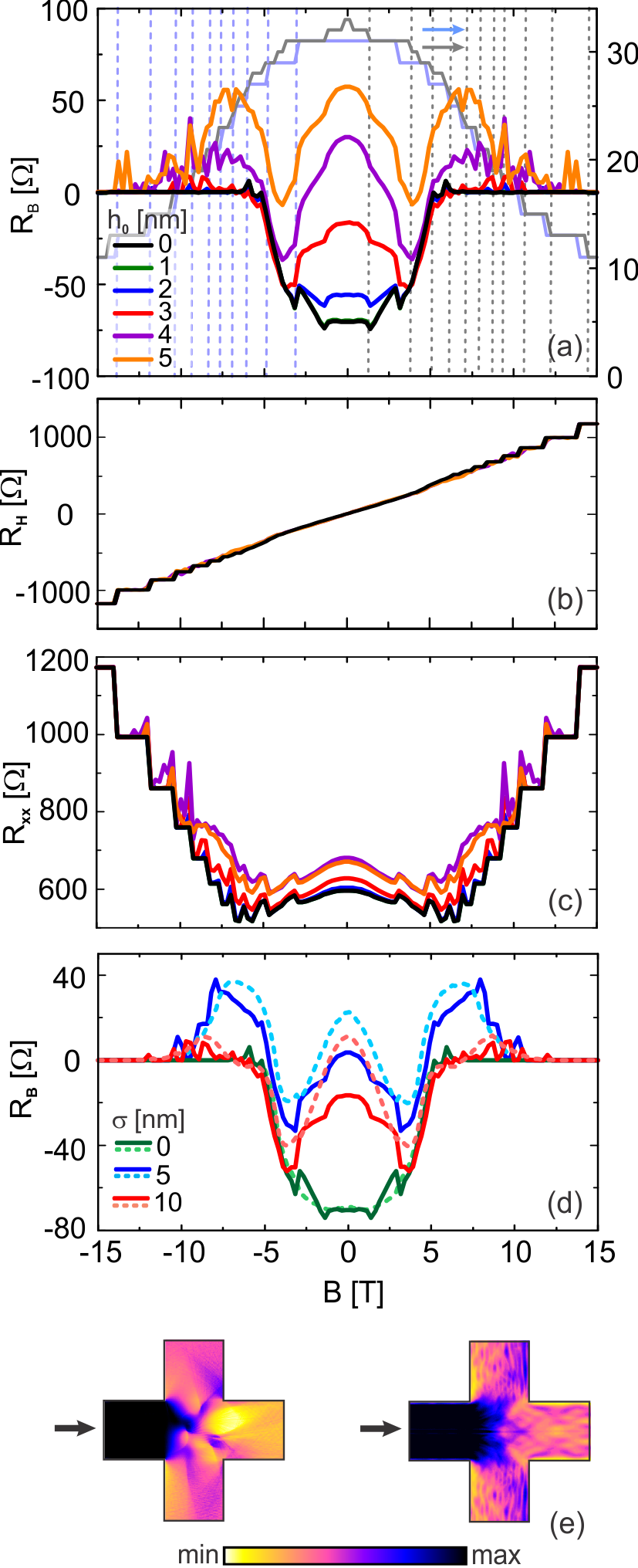}
\caption{(a) The bend, (b) Hall, and (c) longitudinal resistance  versus the applied magnetic field for different values of the height $h_0$ of the bump given in the inset of (a). The bend resistance is shown with the number of conducting channels, $N$, in zigzag (light violet curve) and armchair (gray curve) terminal (right vertical axis). The vertical lines indicate when the number of conduction channels $N$ changes in the zigzag (marked for negative $B)$ and armchair (marked for positive $B$) terminal. Calculations are made using tight-binding model for $E_F = 0.3$ eV, $\sigma = 10$ nm and $L = W = 100$ nm. (d) Comparison of the bend resistances obtained using the tight-binding (solid lines) and classical (dashed lines) method for different values of $\sigma$ shown in the inset. (e) Current density obtained using classical (left) and tight-binding (right) method.}
\label{f7}
\end{center}
\end{figure}

Fig. \ref{f7}(a) shows calculated bend resistance for a few values of the height of the bump given in the inset. Similarly as in Section \ref{sec_cc} a single bump is placed in the center of the system. We see that the general behaviour of the resistance is in agreement with the classical result. The strain causes an increase of the resistance around zero field and the side-peaks occur, as discussed previously. Fig. \ref{f7}(a) shows in addition rather large oscillations of the bend resistance, which are not present in our previous classical result shown in Fig. \ref{f2}(b) and therefore should be of quantum origin. Notice that the positions of those peaks do not change with $h_0$. Thus, their origin is not related to the bump. 

As was mentioned earlier, due to the relatively small size of our device the Fermi energy was increased in order to increase the number of conducting channels, $N$. Nevertheless, this number is still fairly small (at $B = 0$ there are 31 conduction channels in the zigzag and 33 conduction channels in the armchair lead). Furthermore, the bend resistance is low (an order of magnitude lower than the Hall resistance), hence, any fluctuation of $N$ would have a high impact on the bend resistance. Therefore, in Fig. \ref{f7}(a) we add the number of conducting channels, $N$, in the two leads versus the applied magnetic field (the two step-like curves referenced to right axis). The vertical lines show the magnetic field values at which either $N$ for the zigzag (marked only at negative values of $B$) or the armchair (marked only at positive values of $B$) terminal changes. We see that indeed the change of $N$ is correlated with the occurrence of oscillations in the resistance. 

The Hall resistance, shown in Fig. \ref{f7}(b), similarly as in the classical case is almost insensitive to the presence of the bump. For small fields the Hall resistance exhibits a linear dependence on the magnetic field with additional small kinks usually referred to as the last plateau\cite{rce01, rbm2, rhr10}, which is a consequence of ballistic motion of our carriers. On the other hand, at higher magnetic fields the Hall resistance shows quantized steps, a feature that our classical model couldn't capture. 

Fig. \ref{f7}(c) shows the change of the longitudinal resistance, $R_{XX} = R_{13, 31}$, with the magnetic field for a few values of $h_0$. If there is no bump the two terminal resistance has a local maximum around zero magnetic field. Increase of the external magnetic field causes a slight drop in the longitudinal resistance at lower fields and oscillations related to the change of the number of conducting channels occur. At high fields longitudinal resistances become quantized because in a two-terminal measurement the longitudinal and Hall resistance are mixed. The bump causes an increase of $R_{XX}$ at lower fields, similar as for the bend resistance. On the other hand, figure shows less clear quantized steps in the resistance at higher fields due to the interaction of the electrons with the bump. Notice that at very high fields steps in the resistance are unaffected by the bump since the conducting electrons are localized at the edge and do not interact with the strained region.

Comparison of the classical and the tight-binding method is presented in Fig. \ref{f7}(d). Here, we show the bend resistances obtained for two values of $\sigma$ given in the inset, as well as for the case when there is no bump, i.e. $\sigma = 0$. The classical results are shown by the dashed curves while the results of our tight-binding calculations are given by the solid curves. The figure shows rather good agreement between the two methods especially in the case when there is no bump. However, in the case when there is a bump in the system the height of the peak around $B=0$ differs significantly. We found that this difference is due to the transmission probability $T_{31}$ which is higher in the tight-binding case. This is confirmed by the current intensity plots shown in Fig. \ref{f7}(e). We see that in the classical case, shown in the left panel, more electrons are reflected on the bump which leaves a bright spot (low current density) behind the bump. In the tight binding case, on the other hand, more electrons are able to transmit through the bump and hence transmission probability $T_{31}$ increases. Our study shows that the electrons that transmit through the center of the bump all belong to the same valley.

\section{Conclusion}
\label{sec_con}
In this paper we investigated the influence of  strain induced by a Gaussian bump on the magneto-transport in a graphene Hall bar. The simulations were performed using the classical billiard model as well as the quantum transport equation with a nearest-neighbour tight-binding model for graphene. Both models result in similar qualitative results. 

The hallmarks of the effect of a Gaussian bump is the decrease of the negative resistance around zero applied field and the occurrence of side-peaks in the bend resistance. The negative bend resistance is a consequence of ballistic transport in multi-terminal devices. Hence, the decrease of the resistance around $B = 0$ tells us that the bump acts as a scatterer that blocks the flow of electrons towards the opposite terminal. However, at higher fields, for which the cyclotron radius becomes smaller than the half-width of the terminal, the bump enhances the resistance by bending the electron trajectories towards different terminals. The amplitude of the side peak in $R_{B}$ increases with the height of the bump while its position is determined by $2r_{c} \approx W$ independent of $h_0$.

We showed that based on the parameters of the bump we are able to predict the value of the external magnetic field at which the bend resistance drops to zero. This prediction could be written as a simple formula that is based on geometrical arguments. Alternatively, if the external magnetic field is known we can use this formula to estimate the size of the Gaussian bump.

The Hall resistance, on the other hand, showed a weak dependence on the presence of the bump. The reason for this was explained by the fact that regions with opposite directions of the magnetic field would bend electrons to opposite terminals and thus the Hall resistance will remain (almost) unchanged. When the bump is moved out of the center of the Hall cross the Hall resistance is no longer zero for zero applied field and we predict valley polarized electrons in the Hall leads. 

The quantum tight-binding calculations showed similar qualitative features as the classical results. Both, the decrease of resistance around zero applied field and the occurrence of side-peaks, are observed in the bend resistance. However, additional oscillations were present in $R_B$ that could be connected with changes in the number of conducting channels, $N$.

\section{Acknowledgement}
\label{sec_ack}

This work was supported by the Flemish Science Foundation (FWO-Vl) and
the European Science Foundation (ESF) under the EUROCORES Program
EuroGRAPHENE within the project CONGRAN.
\end{document}